\documentclass{PoS}
\usepackage{graphicx}
\usepackage{amsmath, amssymb}
\newcommand{\beq}{\begin{equation}}
\newcommand{\eeq}{\end{equation}}
\newcommand{\bea}{\begin{eqnarray}}
\newcommand{\eea}{\end{eqnarray}}
\newcommand{\nn}{\nonumber\\}

\newcommand{\BAN}{\begin{eqnarray*}}
\newcommand{\EAN}{\end{eqnarray*}}

\def\g5{\gamma_5}
\newcommand{\wideh}{}

\title{One-flavor algorithm for Wilson and domain-wall fermions}

\ShortTitle{One-flavor algorithm for Wilson and domain-wall fermions}

\author{\speaker{Kenji Ogawa}$^{,a}$%
        \thanks{E-mail: ogawak@phys.ntu.edu.tw},
       Ting-Wai~Chiu$^{a,b}$%
       \thanks{E-mail: twchiu@phys.ntu.edu.tw},
       Tung-Han~Hsieh$^c$ (for the TWQCD Collaboration) \\  \\
$^{a}$ Physics Department, and Center for Theoretical Sciences,  
National Taiwan University, Taipei~10617, Taiwan \\
$^{b}$ Center for Quantum Science and Engineering, National Taiwan University, Taipei~10617, Taiwan \\
$^{c}$ Research Center for Applied Sciences, Academia Sinica, Taipei~115, Taiwan
}

\abstract{ 
We construct positive-definite pseudofermion actions for one fermion flavor in lattice field theory, 
for Wilson and domain-wall fermions respectively.
The positive definiteness of these actions ensures that they can be simulated with the  
Hybrid Monte Carlo (HMC) method. 
For lattice QCD with optimal domain-wall quarks, 
we compare the efficiency of HMC simulations of 2-flavor and $(1+1)$-flavor,
and find that the efficiency ratio is about $3:2$. }

\FullConference{The XXVII International Symposium on Lattice Field Theory\\
          July 26-31, 2009\\}
          
\begin{document}

\section{Introduction}

In Monte Carlo simulations of unquenched lattice QCD, 
a positive-definite fermion determinant is required 
to ensure the convergence of the Markov chain. For Hybrid Monte Carlo (HMC) simulation \cite{Duane:1987de}, 
it also requires that the pseudofermion action is positive-definite such that the pseudofermion field 
can be simulated with a Gaussian noise.  
For QCD with two degenerate flavors, this is guaranteed provided that $ \det( D^\dagger) = \det (D) $.
The challenge of simulating one fermion flavor with HMC is that 
it is nontrivial to write down a positive-definite pseudofermion action    
with determinant exactly equal to $ \det(D) $.   
So far, there are several strategies to simulate one fermion flavor with HMC,   
e.g., to approximate $ D^{-1} $ by a polynomial of even degree \cite{Borici:1995am}, 
or to approximate $ (D^\dagger D)^{-1/2} $ by a rational polynomial function \cite{Clark:2006fx}. 
In this paper, we construct a positive-definite pseudofermion action 
with determinant exactly equal to $ \det(D) $,    
for Wilson and domain-wall fermions respectively. 

The outline of this paper is as follows.
In Sec.\,\ref{sec:wilson}, we construct the pseudofermion action for one-flavor Wilson fermion, 
as well as that with the Hasenbusch acceleration \cite{Hasenbusch:2001ne}. 
In Sec.\,\ref{sec:domainwall}, we construct the pseudofermion action for 
one-flavor domain-wall fermion.
In Sec.\,\ref{sec:odwf}, we construct the pseudofermion action for 
one-flavor optimal domain-wall fermion.
In Sec.\,\ref{sec:numerical}, 
we compare the HMC simulations of 2 degenerate flavors with that of one flavor, 
for lattice QCD with optimal domain-wall fermion. 
In Sec.\,\ref{sec:conclusion}, we conclude with some remarks.  

\section{Wilson fermion} \label{sec:wilson}
In this section, we derive a positive-definite pseudofermion action which gives exactly the determinant 
of the Wilson-Dirac operator 
\beq
D_W 
= W + m + \sum_\mu t_\mu \gamma_\mu
=
\begin{pmatrix}
( W + m ) { \bf 1 } & \sum_\mu t_\mu \sigma_\mu \\
\sum_\mu t_\mu \sigma_\mu^\dagger  & ( W + m ) { \bf 1 } 
\end{pmatrix},
\label{eq:wilson}
\eeq
where
\beq
W = 
- \frac{1}{2} \sum_\mu 
\left [ U_{\mu}(x) \delta_{x + \hat \mu, y} + U^\dagger_{\mu}(x - \mu) \delta_{x - \hat \mu, y} \right ] + 4, 
\hspace{2mm}
t_\mu =
\frac{1}{2} 
\left [ U_{\mu}(x) \delta_{x + \hat \mu, y} - U_{\mu}^\dagger(x - \mu) \delta_{x - \hat \mu, y} \right ],
\nonumber
\eeq
\beq
\gamma_\mu = 
\begin{pmatrix}
0 & \sigma_\mu \\
\sigma_\mu^\dagger  & 0  
\end{pmatrix},
~~\sigma_\mu = ( i { \bf 1 }, \sigma_i ),  
\nonumber
\eeq
and $ \sigma_i \, (i = 1, 2, 3 ) $ are Pauli matrices.
Using the Schur decomposition, the determinant of (\ref{eq:wilson}) can be written as
\bea
\label{eq:DWH}
\det D_W = \det(W + m)^2 \det W_H ,  
\eea
where $ W_H $ is the Schur complement of $ D_W $, i.e.,  
\bea
W_H = W + m - \sum_{\mu, \nu} t_\mu \frac{1}{W + m} t_\nu \sigma_\mu^\dagger \sigma_\nu .   
\eea

The positive definiteness of $ W_H $ is asserted as follows. 
For any background gauge field, the eigenvalues of $ W $ and $ (\sigma \cdot t) $ 
satisfy the inequalities:  ${ 0 \le \lambda(W) \le 8 }$, and 
${| \lambda(\sum_\mu \sigma_\mu t_\mu) | \le 4 }$. 
It follows that $ W_H $ is positive-definite for $ m > 4 $. 
Now, if $ m $ is decreased from 4 to a smaller value, 
then the smallest eigenvalue of $ (W + m) $ is also decreased.
Suppose that at $ m = m_{cr} $ (the largest value), it becomes zero, 
then $ \det(W + m_{cr}) = 0 $, and $ W_H(m_{cr}) $ becomes singular, 
thus $ W_H $ is no longer positive-definite.     
From (\ref{eq:DWH}), we see that whenever $ W_H (m) $ has a zero eigenvalue, 
$ D_W (m) $ also has a zero eigenvalue.
On the other hand, when $ W_H (m) $ is singular due to a zero eigenvalue of $ ( W + m ) $, 
$ D_W(m) $ does not necessarily have a zero eigenvalue, which follows from the inequality,  
$\lambda_{min}(W+m) \le \mathrm{Re} (\lambda(D_W(m))) \le \lambda_{max}(W+m)$.
In other words, the largest $ m $ (i.e., $ m_{cr} $) which gives $ \det(W + m) = 0 $ 
is equal to or larger than the largest $ m $ which gives $ \det D_W(m) = 0 $.

Thus the pseudofermion action for one-flavor Wilson fermion can be written as 
\bea
S_{PF} 
= \Phi_1^\dagger (W + m)^{-2} \Phi_1 + \Phi_2^\dagger ( W_H )^{ -1 } \Phi_2  
= \Phi_1^\dagger (W + m)^{-2} \Phi_1 
   - \left ( 0 ~~ \Phi_2^\dagger \right ) H_W^{-1} \begin{pmatrix} 0 \\ \Phi_2 \end{pmatrix}
\label{eq:wls_spf}
\eea
where $H_W = \gamma_5 D_W $, $ \Phi_1 $ is a pseudofermion field without Dirac index, 
and $ \Phi_2 $ is a pseudofermion field with 2 spinor components. 

To generate the pseudofermion field $ \Phi_2 $ from a Gaussian random noise field 
$ \Xi_2 $, we need to take the square root of $ W_H $, i.e., $ \Phi_2 = \sqrt{ W_H } \Xi_2 $ .
Here we use the Zolotarev optimal rational approximation for the square root,  
\beq
\Phi_2 = \sqrt{ W_H } \, \Xi_2
\simeq
\left(p_0 + \sum_{l=1}^{N_\mathrm{app}} 
\frac{p_l}{q_l + W + m - \sum_{\mu, \nu} t_\mu \frac{1}{W + m} t_\nu \sigma_\mu^\dagger \sigma_\nu } \right) \Xi_2 , 
\label{eq:sqrtWH_Xi2}
\eeq
where $ p_0 $, $ p_l $ and $ q_l $ are expressed in terms of Jacobian elliptic functions. 
At first sight, the operations in (\ref{eq:sqrtWH_Xi2}) look formidable.
However, since $ W_H $ is the Schur complement of $ D_W $, each term in (\ref{eq:sqrtWH_Xi2}) 
can be obtained by the inversion of $ (D_W + q_l P_-) $, i.e.,  
\beq
\begin{pmatrix}
0 \\ \Phi_2 
\end{pmatrix}
=
P_- \left( p_0 + \sum_{l=1}^{N_\mathrm{app}} p_l \left ( D_W + q_l P_- \right )^{-1} \right)
\begin{pmatrix}
0 \\ \Xi_2 
\end{pmatrix} .
\label{eq:sqrtWH}
\eeq
Note that one cannot apply the multi-shift conjugate gradient algorithm in (\ref{eq:sqrtWH}), 
since $ P_- $ does not commute with $ D_W $. 
However, these $N_\mathrm{app} $ number of inversions can be speeded up using 
chronological inversion method \cite{Brower:1995vx}.
\subsection{Wilson fermion with Hasenbusch acceleration}
The idea of Hasenbusch's method \cite{Hasenbusch:2001ne} 
is to introduce a heavy pseudofermion field (with mass $ M > m $) such that
the speed of simulating both $ \det( D_W(m)/D_W(M) ) $ and $ \det( D_W(M) ) $ 
is faster than that of $ \det( D_W(m) ) $. To apply Hasenbusch's method, we consider
\bea
\det \left( \frac{D_W(m_1)}{D_W(m_2)} \right)
= \det \left( \frac{ W + m_1 }{W + m_2} \right)^2  
  \det \left( \frac{ W_H(m_1)}{\overline W_H(m_2)} \right)
\label{eq:det_ghost}
\eea
where 
\bea
\overline W_H(m) \equiv 
{W + m} - \sum_{\mu,\nu} t_\mu \frac{1}{W + m} t_\nu \sigma_\mu \sigma_\nu^\dagger. 
\eea
Now, using the Schur decomposition, we can prove that
\beq
\det \left( D_W(0) + m_1 P_+ + m_2 P_- \right)
= \det( W + m_1 )^2 \det[ W_H(m_1) + \Delta_m ]
= \det( W + m_2 )^2 \det[ \overline W_H(m_2) - \Delta_m ], 
\label{eq:det_lr}
\eeq
where $ \Delta_m = m_2 - m_1 > 0 $. 
Using this relation, we can rewrite (\ref{eq:det_ghost}) as
\beq
\det \left ( \frac{D_W(m_1)}{D_W(m_2)} \right )
=
\det \left( \frac{ W_H(m_1)}{ W_H(m_1) + \Delta_m  } \right) 
\det \left( \frac{ \overline W_H(m_2) - \Delta_m  }{ \overline W_H(m_2) } \right ). 
\label{eq:det_hbsch}
\eeq
Next we assert the positive definiteness of both operators on the RHS of (\ref{eq:det_hbsch}).
First, we assume that $ m_2 $ is sufficiently large such that $ (W + m_2 ) $,  
$ W_H (m_2)$, and $ \overline W_H (m_2) $
are positive-definite. Then, the second factor on the RHS of (\ref{eq:det_hbsch}) is 
positive-definite for $ m_1 > m_{1}^{B} $ where $ m_1^{B} $ is the largest value 
satisfying $ \det[ D_W(0) + m_1^{B} P_+ + m_2 P_- ] = 0 $, as a consequence of (\ref{eq:det_lr}).
Now, for the operator $  W_H(m_1)( W_H(m_1) + \Delta_m )^{-1} = ( 1 + \Delta_m / W_H(m_1) )^{-1} $,   
we note that even if $ W_H(m_1) $ is not well-defined, 
$ W_H( m_1)^{-1} = P_- H_W(m_1)^{-1} P_- $ can still be well-defined.
From (\ref{eq:det_hbsch}), $ \det ( W_H(m_1) ( W_H( m_1 ) + \Delta_m )^{-1}) $ is positive-definite 
if both $  \det D_W(m_1) $ and $ \det [ \overline W_H(m_2) - \Delta_m ] $ are positive-definite. 
It follows that both operators on the RHS of (\ref{eq:det_hbsch})
are positive-definite for $ m_1 > \max( m_1^A, m_1^B) $, 
where $ m_1^B $ has defined above, 
and $ m_1^{A} $ is the largest value satisfying $ \det D_W(m_1^{A}) = 0 $. 

Thus the pseudofermion action for one-flavor Wilson fermion with Hasenbusch acceleration 
can be written as
\bea
&& S_{PF} 
=  
  \Phi_3^\dagger \left ( 1 + \frac{ \Delta_m }{ W_H(m_1) } \right ) \Phi_3
+ \Phi_4^\dagger \left ( 1 + \frac{ \Delta_m }{ [ \overline W_H(m_2) - \Delta_m ] } \right ) \Phi_4 \nonumber \\
&=& \Phi_3^\dagger \Phi_3 + \Phi_4^\dagger \Phi_4 
- \Delta_m \left ( 0 ~~ \Phi_3^\dagger \right )
H_W^{-1} ( m_1 ) \begin{pmatrix} 0 \\ \Phi_3 \end{pmatrix}
+ \Delta_m \left ( \Phi_4^\dagger ~~ 0 \right )
\left ( H_W(0) + m_1 P_+ - m_2 P_- \right )^{-1}
\begin{pmatrix} \Phi_4 \\  0 \end{pmatrix}, 
\nonumber 
\label{eq:WH_SPF}
\eea
where $ \Phi_3 $ and $ \Phi_4 $ are pseudofermion fields with 2 spinor components.

\section{Domain-wall fermion}\label{sec:domainwall}
The basic idea of domain-wall fermion is to use $ N_s $ layers of 4-d Wilson-Dirac fermions 
(with nearest neighbor coupling in the 5-th dimension) such that 
the exactly chiral fermion fields emerge at the boundary layers 
in the limit $ N_s \to \infty $ \cite{Kaplan:1992bt}.
The domain-wall fermion operator can be defined as  
\bea
  {\mathcal D}_\mathrm{dwf}(m) & = & W - m_0 + \sum_\mu \gamma_\mu t_\mu + \wideh M(m)
\\ 
\wideh M(m) & = & P_+ \wideh M_+(m) + P_- \wideh M_-(m)
\eea
\beq
\wideh M_+(m)_{s,s'} =
\begin{cases} 
\delta_{s',s} - \delta_{s',s - 1},  &  1 < s \le N_s,   \\
\delta_{s',s} + m \delta_{s', N_s}, &  s = 1, 
\end{cases}
\hspace{4mm}
\wideh M_-(m)_{s,s'} =
\begin{cases} 
\delta_{s',s} - \delta_{s',s + 1}, & 1 \le s < N_s,  \\
\delta_{s',s} + m \delta_{s', 1},  & s = N_s,  
\end{cases}
\eeq
where $ m $ is the fermion mass, and $ m_0 \in (0,2) $ is a parameter called "domain-wall height".
Using the Schur decomposition, the determinant of domain-wall fermion operator can be written as
\beq
 \det {\mathcal D}_\mathrm{dwf}(m)
 =  
\det \left [ W - m_0 + \wideh M_+(m) \right ]^2
\det {\mathcal W}_H(m)
 =  
\det \left [ W - m_0 + \wideh M_-(m) \right ]^2
\det \, \overline {\mathcal W}_H(m) , 
\eeq
where 
\beq
{\mathcal W}_H(m)
=
R_5 \left ( W - m_0 + \wideh M_-(m)
- t_\mu \frac{ 1 }{ W - m_0 + \wideh M_+(m) } t_\nu \sigma_\mu^\dagger \sigma_\nu \right )
\eeq
\beq
\overline {\mathcal W}_H(m)
=
R_5 \left ( W - m_0 + \wideh M_+(m)
- t_\mu \frac{ 1 }{ W - m_0 + \wideh M_-(m) } t_\nu \sigma_\mu \sigma_\nu^\dagger \right ).
\eeq
Here $ R_5 $ is the reflection operator in the 5-th dimension,  
$ \left( R_5 \right)_{s, s'}  = \delta_{s, N_s + s' - 1}$,  
which is introduced such that $ {\mathcal W}_H(m) $ and $ \overline {\mathcal W}_H(m) $ are hermitian.
After incorporating the contribution of the Pauli-Villars fields, 
the fermion determinant of domain-wall fermion becomes
\beq
\frac { \det {\mathcal D}_\mathrm{dwf}(m) }
{ \det {\mathcal D}_\mathrm{dwf}(1) }
 = 
\frac { 
\det \left[ W - m_0 + \wideh M_+(m) \right]^2
\det {\mathcal W}_H(m) }{
\det \left[ W - m_0 + \wideh M_+(1) \right]^2
\det \, \overline {\mathcal W}_H(1) 
}
\label{eq:dwf_fdet}
\eeq
Using the Schur decomposition of 
$ \left[ {\mathcal D}_\mathrm{dwf}(1) - \left ( \wideh M_+(1) - \wideh M_+(m) \right ) P_+ \right] $,
we obtain the relation
\bea
\det \left [ W - m_0 + \wideh M_+(m) \right ]^2
\cdot 
\det \left [ {\mathcal W}_H(m) + \wideh \Delta_-(m) \right ] =
\det \left [ W - m_0 + \wideh M_-(1) \right ]^2
\cdot 
\det \left [ \overline {\mathcal W}_H(1) - \wideh \Delta_+(m) \right ], \nn 
\label{eq:det_lr1}
\eea
where
\bea
\label{eq:delta_m}
\left [ \wideh \Delta_+(m) \right ]_{s,s'} 
= (1 - m) \, \delta_{s, N_s } \, \delta_{ s', N_s },  \hspace{4mm} 
\left [ \wideh \Delta_-(m) \right ]_{s,s'} 
=  (1 - m) \, \delta_{s,1} \, \delta_{s',1} .
\eea
Using (\ref{eq:det_lr1}), we can rewrite (\ref{eq:dwf_fdet}) as 
\bea
\det \left [
{\mathcal W}_H(m)
/ \left ( 
{\mathcal W}_H(m) + \wideh \Delta_-(m)
\right )
\right ]
\cdot
\det \left [
\left (
\overline {\mathcal W}_H(1) - \wideh \Delta_+(m)
\right )
/
\overline {\mathcal W}_H(1)
\right ], 
\nonumber 
\eea
and its inverse 
\beq
\det \left [
1 +
\wideh \Delta_-(m)
\frac{1}{ {\mathcal W}_H(m) }
\right ]
\cdot
\det \left [
1 + 
\wideh \Delta_+(m)
\frac{1}{
\overline {\mathcal W}_H(1) - \wideh \Delta_+(m) }
\right ], 
\label{eq:det_dwf}
\eeq
can be used to construct the pseudofermion action.
Using (\ref{eq:delta_m}), we can simplify (\ref{eq:det_dwf}) to
\beq
\det \left(1+(1-m) \left[\frac{1}{{\mathcal W}_H(m)} \right]_{s = s' = 1} \right ) 
\cdot
\det \left(1+(1-m) \left[\frac{1}{\overline {\mathcal W}_H(1)-\wideh \Delta_+(m)} \right]_{s = s' = N_s} \right ).  
\label{eq:dwf_det}
\eeq 
Thus we can write the pseudofermion action for one-flavor domain-wall fermion as 
\bea
S_{PF} 
&=& \Phi_1^\dagger \Phi_1 - 
(1 - m)\,
( 0 ~~ \Phi_1^\dagger )
\left[ \gamma_5 R_5 {\mathcal D}_\mathrm{dwf}(m) \right]^{-1}_{s = s' = 1}
\begin{pmatrix}
0 \\ \Phi_1 
\end{pmatrix}
\nonumber \\
&+&  \Phi_2^\dagger \Phi_2
+
(1 - m) \,
( \Phi_2^\dagger ~~  0 )
\left [ \gamma_5 R_5 {\mathcal D}_\mathrm{dwf}(1) - \wideh \Delta_+(m) P_+ \right ]^{-1}_{s = s' = N_s} 
\begin{pmatrix}
\Phi_2 \\ 0 
\end{pmatrix}
\nonumber 
\eea
where $ \Phi_1 $ and $ \Phi_2 $ are pseudofermion fields (on the 4-dimensional lattice) with 2 spinor components.
%
%
Now we assert that the operators in (\ref{eq:dwf_det}) are positive-definite for $ m > 0 $. 
At $ m = 1 $, they are equal to the identity operator, thus are positive-definite.
As $ m $ is decreased to be less than one, their positive definiteness will lose only
if either of the determinants in (\ref{eq:dwf_det})
becomes zero or singular. 
Using $ [W, M_-(1)] = [W,M_+(m)] = 0 $, and the fact that   
the eigenvalues of $M_-(1)$ and $M_+(m)$ have non-zero imaginary parts,  
we immediately see that ${ ( W - m_0 + \wideh M_-(1) ) }$ and ${ ( W - m_0 + \wideh M_+(m) ) }$
cannot have zero eigenvalue for $ m > 0 $.
Thus the operators in (\ref{eq:dwf_det}) are well-defined for $ m > 0 $.
Furthermore, since (\ref{eq:det_dwf}) is equal to the determinant of the massive overlap-Dirac operator
with polar approximation for the sign function of $ H = \gamma_5 D_w ( 2 + D_w)^{-1} $, 
thus for $ m > 0 $, it must be positive and it follows that 
the operators in (\ref{eq:dwf_det}) are positive-definite. 
\section{Optimal domain-wall fermion}\label{sec:odwf}
For finite $ N_s $, the chiral symmetry of the domain-wall fermion is not exact. 
However, one can attain the (mathematically) maximal chiral symmetry by assigning 
each layer in the 5-th dimension a different weight $ \omega_s $, 
according to the formula derived in Ref. \cite{Chiu:2002ir}. Then the effective 4D
Dirac operator is exactly equal to the overlap-Dirac operator with Zolotarev optimal rational 
approximation to the sign function of $ H_w $. 
The optimal domain-wall fermion operator \cite{Chiu:2002ir} is defined as 
\beq
{\mathcal D}_{opt}(m) = { \boldsymbol \omega } D_W(-m_0) \left ( 1 + \wideh L(m) \right )
+ \left ( 1 - \wideh L(m) \right ), 
\hspace{4mm} \wideh L(m) = P_+ \wideh L_+(m) + P_- \wideh L_-(m),   
\eeq
where
\beq	
\wideh L_+(m)_{s,s'} =
\begin{cases} 
  \delta_{s',s - 1},  &  1 < s \le N_s,    \\
 - m \delta_{s', N_s}, &  s = 1, 
\end{cases}
, \hspace{4mm}
\wideh L_-(m)_{s,s'} =
\begin{cases} 
 \delta_{s',s + 1}, &  1 \le s < N_s, \\
 - m \delta_{s', 1},  &  s = N_s. 
\end{cases}
\eeq
Here $\boldsymbol \omega $ denotes the diagonal matrix of the weights $ \{ \omega_s, s=1, \cdots, N_s \} $.
Moreover, we can construct $ \boldsymbol \omega $ such that it satisfies 
$ R_5 \boldsymbol \omega = \boldsymbol \omega R_5 $. 
To derive a positive-definite pseudofermion action for one flavor, we can start from  
\beq
{\mathcal D}'_{opt}(m) = {\boldsymbol \omega} D_W(-m_0) + ( 1 - \wideh L(m) )/( 1 + \wideh  L(m) )
\eeq
which is different from $ {\mathcal D}_{opt}(m) $ by a matrix factor independent of the gauge field.  
Using the same technique we have used for the case of domain-wall fermion, we obtain the positive-definite 
pseudofermion action as follows\footnote{Detailed derivation will be given in a forthcoming paper.}
\bea
S_{PF}
& = &
  \Phi_1^\dagger \Phi_1 
+ \Phi_2^\dagger \Phi_2
- \left( \frac{1 - m}{ 1 + m } \right )
(0~~~(-1)^{s} \Phi_1^{\dagger}) 
\left[ \gamma_5 R_5 \mathcal D_{opt}'( m ) \right]^{-1}_{ s, s'}
\begin{pmatrix} 0 \\ (-1)^{s'} \Phi_1 \end{pmatrix}
\nonumber
\\ 
&& \hspace{-18mm} +
\left( \frac{1 - m}{ 1 + m } \right )
((-1)^{s} \Phi_2^{\dagger}~~~0) 
\left [ 
\gamma_5 R_5 \mathcal D_{opt}'( 1 ) 
- R_5 \left( \frac{1 - \wideh L(1)}{1 + \wideh L(1)} - \frac{1 - \wideh L(m)}{1 + \wideh L(m)} \right ) P_+ 
\right ]^{-1}_{ s, s'}
\begin{pmatrix} (-1)^{s'} \Phi_2 \\ 0 \end{pmatrix}
\label{eq:ODWF_1f}
\eea
where $ \Phi_1 $ and $ \Phi_2 $ are pseudofermion fields with 2 spinor components on the 4-dimensional lattice.

\section{Numerical test}\label{sec:numerical}

We compare the efficiency of HMC simulation of 2-flavor and $ (1+1) $-flavor QCD with 
optimal domain-wall quarks, 
on the $ 12^3 \times 24 \times 16 (N_s) $ lattice.  
For the gluon part, we use Iwasaki gauge action at $ \beta = 2.30 $.
In the molecular dynamics, we use the Omelyan integrator \cite{Takaishi:2005tz}, 
and the Sexton-Weingarten multiple-time scale method \cite{Sexton:1992nu}.
The time step for the gauge field ($\Delta \tau_\mathrm{Gauge}$) is the same 
for both 2-flavor and $(1+1)$-flavor cases, while 
the time step ($\Delta \tau_\mathrm{PF}$) for the pseudofermion field 
in the $(1 + 1)$-flavor case is 4 times larger than that for the 2-flavor case 
such that the acceptance rate is roughly the same for both cases. 
We use conjugate gradient (CG) with mixed precision for the inversion of 
the quark matrix (with even-odd preconditioning). The length of each trajectory is set to 2.
After discarding 300 trajectories for thermalization, we accumulate 100 trajectories for 
the comparison of efficiency. Our results are given in Table.\ref{table:test1}.
We see that acceptance rate is almost the same for $(1+1)$-flavor and 2-flavor simulations. 
If auto-correlation time is the same, then the efficiency of HMC can be estimated by 
the total acceptance divided by the CG iteration number, 
and the efficiency ratio for 2-flavor and $(1+1)$-flavor is about $ 3:2$. 

\begin{table}
\begin{center}
\begin{tabular}{|c|c|c|c|c|c|c|} \hline
$ m $ &
$N_\mathrm{f}$ &
$N_\mathrm{Iter}^\mathrm{(HB)} / 10^{3} $ &
$N_\mathrm{Iter}^\mathrm{(MD)} / 10^{3} $ &
$N_\mathrm{Iter}^\mathrm{(Total)} / 10^{3} $ & Acceptance &
$ \mathrm{Acceptance}/N_\mathrm{Iter}^\mathrm{(Total)}$  \\ \hline
$0.019$ & $1 + 1 $ &  75\,(1) & 260\,(1) & 345\,(1) & 0.88\,(3) & $ 2.6\,(1) \times 10^{-6} $ \\ \cline{2-7}
        & $ 2      $ & 0.6\,(1) & 239\,(2) & 240\,(2) & 0.90\,(3) & $ 3.8\,(2) \times 10^{-6} $ \\ \hline
$0.038$ & $1 + 1 $ &  54\,(1) & 125\,(1) & 179\,(1) & 0.90\,(3) & $ 5.0\,(2) \times 10^{-6} $ \\ \cline{2-7}
        & $2       $ & 0.6\,(1) & 112\,(1) & 113\,(1) & 0.91\,(3) & $ 8.0\,(3) \times 10^{-6} $ \\ \hline
\end{tabular}
\caption{Comparison of HMC efficiency for the 2-flavor and $(1+1)$-flavor QCD with optimal domain-wall quarks. 
The step size for the gauge field 
$\Delta \tau_\mathrm{Gauge}$  
is $ 0.007 (0.010) $ for $ m = 0.019 (0.038) $. 
while the step size 
$\Delta \tau_\mathrm{PF}$ 
for $ (1+1) $-flavor pseudofermions is 
$ 0.14 (0.20) $ for $ m = 0.019 (0.038) $, 
which is 4 times larger than that for the 2-flavor case. 
Here, $N_\mathrm{Iter}^\mathrm{(HB)}$, $N_\mathrm{Iter}^\mathrm{(MD)}$, and $N_\mathrm{Iter}^\mathrm{(Total)}$
are the average CG iterations for one trajectory (for generating initial pseudofermion fields, molecular dynamics,
and their sum respectively).
}
\label{table:test1}
\end{center}
\vspace{-4mm}
\end{table}
\section{Concluding remark}\label{sec:conclusion}
\vspace{-2mm}
We have constructed a positive-definite pseudofermion action for one fermion flavor in lattice gauge theory, 
with determinant exactly equal to that of the lattice Dirac operator, 
for the Wilson-Dirac operator, and the (optimal) domain-wall fermion operator respectively.
Evidently, these one-flavor positive-definite actions are useful for the HMC simulations 
in nonperturbative studies of the Standard Model (SM), as well as those beyond the SM.  
Our numerical tests show that the step size 
$\Delta \tau_\mathrm{PF}$ (for the pseudofermion field in the molecular dynamics) 
for the ${(1+1)}$-flavor QCD can be 4 times larger than that for the 2-flavor case, 
yet with the same acceptance rate. 
The efficiency ratio for HMC with 2-flavor and $(1+1)$-flavor is about $ 3:2 $. 
With the positive-definite pseudofermion action (\ref{eq:ODWF_1f}) for the strange quark,
TWQCD Collaboration is performing the HMC simulation of (2+1)-flavor QCD
on the $ 16^3 \times 32 \times 16 $ lattice, using a GPU cluster 
\cite{Chiu:2009wh}.
\vspace{-2mm}
\section*{Acknowledgment}
\vspace{-2mm}
   This work is supported in part by
   the National Science Council
   (Nos. NSC96-2112-M-002-020-MY3,
         NSC96-2112-M-001-017-MY3,
         NSC98-2119-M-002-001),
   and NTU-CQSE~(Nos. 98R0066-65, 98R0066-69).

\end{document}